\documentclass[twocolumn,showpacs,showkeys,superscriptaddress]{revtex4}

\usepackage{amsmath}
\usepackage{bbold}
\usepackage{amsfonts} 
\usepackage{amssymb}
\usepackage{pbsi}
\usepackage[T1]{fontenc}
\usepackage{hyperref}
\usepackage{xcolor}
\usepackage{graphicx}
\usepackage{subfig}
\usepackage{float}

\usepackage{color}

\begin{document}

\title{Thermal-Inertial Effects on Relativistic Asymmetric Magnetic Reconnection}

\author{Maricarmen A. Winkler}
\email{ma.winkler822@gmail.com}
\affiliation{Departamento de F\'isica, Facultad de Ciencias, Universidad de Chile, Casilla 653, Santiago, Chile.}

\author{Felipe A. Asenjo}
\affiliation{Facultad de Ingenier\'ia y Ciencias,
Universidad Adolfo Ib\'a\~nez, Santiago 7491169, Chile.}

\begin{abstract}
A general description for relativistic magnetic reconnection is given in terms of asymmetric inflow plasma conditions, such as plasma density, velocity  magnetic field strength, and their respective thermal-inertial effects. In this work, we show that thermal-inertial effects increase  reconnection rates, generalizing previous results. Furthermore, different reconnection rates can be defined depending on the initial asymmetric conditions for the inflow plasma into the reconnection region. We explore two possibilities for reconnection rates, which, depending on the asymmetric magnetization, display different strengths. We discuss the importance on defining   appropriate reconnection rates for   corresponding asymmetric plasma systems in the context of previous results.
\end{abstract}


\maketitle

\section{Introduction}

Magnetic fields are ubiquitous in different occurrences in the universe. From laboratory plasmas to solar wind, flares, magnetospheres, intergalactic space and astrophysical environments, magnetic field lines are present and their interactions are responsible for many of the processes that we are able to observe and study \cite{cveji2022,grasso2001,ryu2008,Ryu2012,Burlaga2001}. Magnetic reconnection is an elemental phenomenon in highly conducting plasmas, where a varying magnetic field can affect the way charged particles move and vice versa. Under certain conditions, magnetic field lines carried along a charged fluid rearrange: they break and merge with other field lines, altering the configuration of a plasma \cite{book_priest2000magnetic,book_biskamp2000magnetic,book_gonzalez2016magnetic}. This process converts magnetic energy into heat and kinetic energy, powering several phenomena in space, astrophysical and laboratory plasmas such as solar wind, evolution of magnetospheres in stars, accretion disks and fusion plasmas, to name a few \cite{zweibel2009,mckinney2011,Zhang2011,Sironi2021}.

Magnetic reconnection has been the subject of many studies, whether theoretical work or simulations, where slow and fast reconnection is treated. Scaling equations and reconnection rates are analyzed to further understand this process, yet most of these studies consider nonrelativistic plasmas \cite{ni2018,Uzdensky2011,cassak2009,Kadowaki2021,Yang2020}. It has been shown over the years that in magnetically dominated environments, relativistic effects have to be considered since the magnetic energy density surpasses the rest mass energy density, meaning that the Alfvén speed of the wave approximates the speed of light \cite{bessho2012fast,takahashi2011scaling,blackman1994}. Relativistic reconnection can be seen pulsar winds, jets from active galactic nuclei and pulsar magnetospheres \cite{lyutikov2003explosive,uzdensky2013physical,mckinney2012reconnection}. 

These previous works have also been developed under a resistive and relativistic magnetohydrodynamical model (resistive RMHD), but it has been shown that thermal-inertial effects can be considered to further enhance this reconnection process \cite{asenjocomisso2014}.
Studies have also mostly included the analysis of a symmetrical scenario, where the magnetic field strength, density, inflow and outflow plasma velocities, and temperatures are all equal for the two reconnecting plasmas. However, it has been frequently observed that different inflow conditions can occur in the heliosphere and near-Earth systems \cite{samadi2017,swisdak2003diamagnetic,malakit2010scaling}, raising the question whether this can occur in other environments, including under relativistic conditions.

Thus in this work we extend previous relativistic reconnection models \cite{mbarek2022}, considering thermal-inertial effects under asymmetrical inflow conditions for the reconnecting plasma. We use a one-fluid generalized relativistic magnetohydrodynamical model to derive scaling equations to study magnetic reconnection in the Sweet-Parker configuration and to obtain the reconnection rate for this process. We find that it is possible to define two different reconnection rates, depending on how the asymmetric conditions of the relativistic inflow plasma are considered. Both of them can be enhanced due to the thermal-inertial effects of the plasma.

The paper is structured as follows. In Sec.~\ref{RMHD} the model equations for a relativistic plasma and basic assumptions are presented. Then, in Sec.~\ref{SPModel} the general asymmetrical and relativistic reconnection model is presented and two different reconnection rates are obtained from this Sweet-Parker configuration, including thermal-inertial corrections. This results are summarized and discussed in Sec.~\ref{discussion}.


\section{Generalized RMHD equations}\label{RMHD}

To estimate the magnetic reconnection rate in the general case of asymmetric inflow conditions in a relativistic plasma when thermal-inertial effects are considered, we use a one-fluid model based on a relativistic two-fluid approximation of pair plasma and electron-ion plasma, proposed by Koide \cite{koide2009}. In this case, relativistic magnetohydrodynamic equations (RMHD) can be used for a plasma with density $\rho$, number density $n$, four-velocity $U_\mu$ (satisfying $U_\mu U^\mu =\eta_{\mu\nu} U^\mu U^\nu=-c^2$, where $c$ is the speed of light), four-current $J^\mu$, and metric signature $\eta_{\mu\nu}=\left(-1,1,1,1\right)$. In this way, the continuity equation is written as 
\begin{align}
    \partial_\mu &\left(\rho U^\mu\right)= 0 \label{eq_cont} \ , 
\end{align}
the generalized momentum equation
\begin{align}
     \partial_\nu \Bigg[\frac{h}{c^2} U^\mu U^\nu + \frac{h}{4n^2e^2c^2}&J^\nu J^\mu\Bigg]=-\partial^\mu p +\frac{1}{c}J_\nu F^{\mu\nu} \label{eq_momentum} \ , 
\end{align}
and the generalized Ohm's law
\begin{align}
      \frac{1}{4ne}\partial_\nu \Bigg[\frac{h}{nec} \big(U^\mu J^\nu +&J^\mu U^\nu\big)\Bigg] \nonumber\\
    =U_\nu F^{\mu\nu}-\eta c \Bigg[J^\mu&+\frac{1}{c^2}U_\alpha J^\alpha U^\mu \left(1+\Theta\right)\Bigg] \label{eq_ohm} \ . 
\end{align} 
In this set of equations, $h$ is the enthalpy of this relativistic plasma, $p$ is the pressure, $e$ the electron charge and $F^{\mu\nu}$ the electromagnetic tensor. Also, $\eta$ is the resistivity and $\Theta$ is the thermal energy exchange rate from negative to positive charged fluids \cite{koide2009}. From Eqs. \eqref{eq_momentum} and \eqref{eq_ohm} it can be seen that thermal inertial effects  appear with terms proportional to the enthalpy. In the momentum equation is associated to the inertia of current density, whereas in Ohm's law the thermal inertial effect correspond to the thermal electromotive force. A thermal function can be defined, depending only on the temperature $T$, such as $f=f\left(T\right)=h/\left(\rho c^2\right)\geq 1$ for all temperature \cite{mahajan2003}.
Finally, along with the previous expressions, it is necessary to complement the plasma system with Maxwell's equations
\begin{equation}
    \partial_\nu F^{\mu\nu}=4\pi J^\mu\ , \hspace{0.8cm} \partial_\nu F^{*\mu\nu}=0\ , \label{eq_maxwell}
\end{equation}
where $F^{*\mu\nu}$ is the dual electromagnetic tensor.

The above model relies in the fact that  physical quantities are defined 
by projecting them in the foliations of spacetime. For example, the electric and magnetic fields can be defined, respectively, as $E^\mu=n_\nu F^{\mu\nu}$ and $B^\mu=n_\nu F^{*\nu\mu}$, where $n^\mu=(1,0,0,0)$. Thus, the electric and magnetic field are defined by projections onto timelike hypersurfaces \cite{asenjocomisso2017}. Similarly, the four-velocity can be put in the form $U^\mu=\gamma c n^\mu+\gamma v^\mu$, such that Lorentz factor is defined as $\gamma=-n_\mu U^\mu/c=(1-v_i v^i/c^2)^{-1/2}$, where we have considered the spacelike four velocity $v^\mu=(0,v^i)$, with
$v^i$ the spatial components of the vectorial plasma fluid  velocity \cite{asenjocomisso2017}.



\section{Asymmetric reconnection model} \label{SPModel}

\begin{figure}[h]\centering
\includegraphics[width=8.5cm]{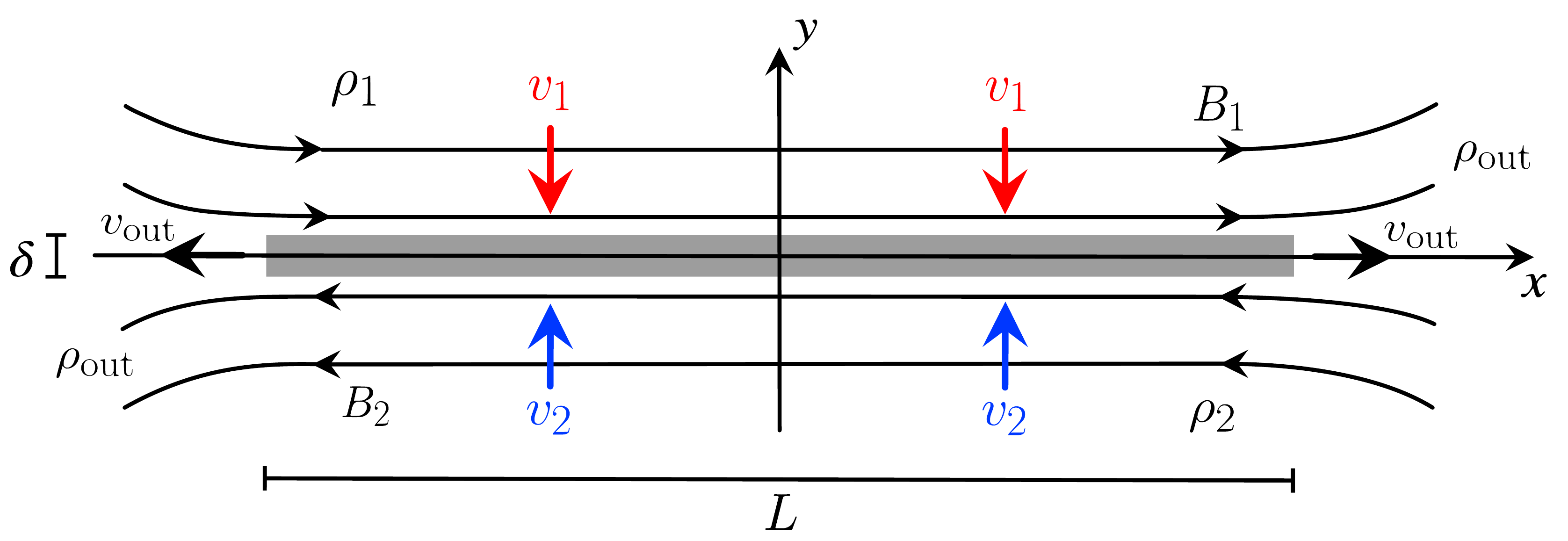}
\caption{Sweet-Parker reconnection configuration for a relativistic asymmetric plasma. The shaded region represents the magnetic diffusion area of length $L$ and width $\delta$.}\label{fig1}
\end{figure}

To estimate the magnetic reconnection rate with asymmetric inflow conditions in the Sweet-Parker configuration, we consider an asymmetric diffusion region of length $L$ and width $\delta$ (with $\delta\ll L$), as shown in Fig. \ref{fig1}.

When the magnetic  reconnection process occurs at the steady state, the diffusion region has two different plasmas flowing in from above and below  to this region (in the $x$ axis). Each plasma flows with its own velocity, $v_1$ from above and $v_2$ from below, into the region of length $L$. Also, each plasma has its own density, $\rho_1$ and $\rho_2$, and magnetic field strength, $B_1$ and $B_2$ (where from now on, subindexes 1 and 2 stand for the above and below regions to the magnetic diffusion region, respectively). Outside of the diffusion region, the plasma can be considered as ideal. 

After the magnetic reconnection takes place, the merged plasma is ejected from the diffusion region with outflow
velocity $v_{\text{out}}$ and density $\rho_{\text{out}}$. In a steady state configuration, a relation between the inflow asymmetric velocities and the outflow velocities can 
be obtained thorugh the continuity equation for the plasma. From Eq.~\eqref{eq_cont} we can estimate the rates
for the conservation of flow as
\begin{equation}
    L\left(\rho_1\gamma_1 v_1+\rho_2\gamma_2 v_2\right)\sim 2\rho_{\text{out}}\gamma_{\text{out}}v_{\text{out}}\delta\ , \label{scale_cont}
\end{equation}
where $\gamma_1$, $\gamma_2$ and $\gamma_{\text{out}}$ are, respectively the Lorentz factors for plasma inflows 1 and 2, and  the plasma outflow.

Ampère's law, $\nabla\times B= 4\pi J /c$ from Eq. \eqref{eq_maxwell}, is written for both inflow plasmas, allowing us to calculate the estimated contribution per region of the plasma current. In this way, equation $-\partial_y B_{1,2}=4\pi J^{z}_{1,2}/c$ , can be used to estimate the current density in the $z$ direction as
\begin{align}
    J^{z}_{1,2}&\sim -\frac{B_{1,2}c}{4\pi \delta}\, . \label{current_density} 
\end{align}
Additionally, the outflow magnetic field strength (in the $y$-direction) can be estimated from magnetic flux conservation for each inflow plasma as $B^y_{1,2}\sim \delta B_{1,2}/L$. Besides, the pressure balance across the layer, from the two different inflow plasmas, becomes $p=\left(B_1^2+B_2^2\right)/8\pi$, since we are considering that  plasmas are magnetically dominant environments far in the incoming region.

On the other hand, along the neutral line there is no contribution from the thermal energy exchange rate between the charged fluids. This due to that in that line $E^x\approx E^y\approx 0$ and $B^z\approx 0,$ while $J^0=0\approx J^x$ and $v^y\approx v^z\approx 0,$ impliying that $U_\mu J^\mu \approx 0$ in this zone. Thus, we can evaluate
the momentum equation \eqref{eq_momentum} along the neutral line, obtaining
\begin{align}
    \partial_x\left(\frac{h_{\text{out}}}{c^2}\gamma_{\text{out}}^2 v^2_{\text{out}}\right) &= -\partial_x p +\frac{1}{c} J_\nu F^{x\nu}\, . \label{approx-momentum}
\end{align}
The  current density inertia can be neglected in the momentum equation due to the direction of $J^\mu$ in the reconnection layer. Then, from Eq. \eqref{current_density}, the outflow magnetic strength flux conservation and the pressure balance, we estimate from Eq. \eqref{approx-momentum} the steady-state relation
\begin{align}
    \frac{h_{\text{out}}}{c^2}\gamma_{\text{out}}^2 v^2_{\text{out}}+p &= -\frac{L}{c} \left(J^z_1 B^y_1-J^z_2 B^y_2\right)\nonumber \\ 
    &\sim \frac{B^2_1+B^2_2}{4\pi} =2p \ . \label{approx-momentum2}
\end{align}
Eq. \eqref{approx-momentum2} shows that $h_{\text{out}}\gamma_{\text{out}}^2 v^2_{\text{out}}/c^2\sim p$. This is the same result shown in Refs.~\cite{asenjocomisso2014, Lyubarsky2005}. Then, by defining the magnetization as $\sigma\equiv B^2/4\pi \rho c^2$, and considering $h_{\text{out}}=f_{\text{out}}\rho_{\text{out}}c^2$, we obtain
\begin{align}
   \gamma^2_{\text{out}} &\sim 1+\frac{\sigma_1\rho_1+\sigma_2\rho_2}{2 \rho_{\text{out}}f_{\text{out}}} \ . \label{gamma}
\end{align}
This estimation for the outflow Lorentz factor
is in agreement with previous results in the symmetric cold limit. In fact, when $\rho_1=\rho_2=\rho_{\text{out}}$,  $\sigma_1=\sigma_2=\sigma$, and $f_{\text{out}}=1$, then we get $\gamma_{\text{out}}=\sqrt{1+\sigma}\approx\sqrt{\sigma}$, in the higly magnetized case $\sigma\gg 1$, case studied by Lyubarsky \cite{Lyubarsky2005}. 

In order to calculate the reconnection rate, we need to estimate the changes introduced by Ohm's law \eqref{eq_ohm}.
As previously mentioned, there is no contribution from the thermal energy exchange rate between the fluids. In this way, in the diffusion region, the generalized Ohm's law is reduced to
\begin{align}
    \frac{1}{4ne}\partial_\nu \left[\frac{h}{ne}\left(U^\mu J^\nu + J^\mu U^\nu\right)\right]&= c\, U_\nu F^{\mu\nu}-\eta J^\mu \ . \label{eq-ohmg}
\end{align}
From Eq.~\eqref{eq-ohmg}, the thermal electromotive effects can be approximated in the reconnection layer as $\partial_\nu \left[h_\text{out}\left(U^\mu J^\nu + J^\mu U^\nu\right)\right]\approx \partial_x \left[h_\text{out} J^\mu U^x\right]\sim h_\text{out}\gamma_\text{out}v_\text{out}J^\mu/L$. Then, Eq. \eqref{eq-ohmg} can be rewritten in the form
\begin{align}
    \left[\frac{h_\text{out}\gamma_\text{out}v_\text{out}}{4 n^2 e^2 L}+\eta \right]J^\mu &=c\,  U_\nu F^{\mu\nu}\, , \label{eq-ohmg2}
\end{align}
where the thermal inertial effects appear in the first term in the right-hand side of Eq. \eqref{eq-ohmg2}. Notice that this term can be put in the form $h_\text{out}\gamma_\text{out}v_\text{out}/\left(4 n^2 e^2 L\right)=f_\text{out} \rho_\text{out}c^2 \gamma_\text{out}v_\text{out}/\left(4 n^2 e^2 L\right)$. Thus, the Ohm's law  yields $J^y=0$, and also
\begin{align}
    \left[\pi f_\text{out}\gamma_\text{out}v_\text{out}\frac{\lambda^2_e}{L}+\eta \right]J^z &\sim c^2 \gamma_\text{out} E^z\ . \label{eq-ohmg3}
\end{align}
where $\lambda_e=c/\omega_p$ is the electron skin depth.

On the contrary, in the ideal region, above and below the reconnection layer, Ohm's law is ideal
\begin{align}
    U_\nu F^{\mu\nu}=0\, , \label{eq-ohmg2ideal}
\end{align}
which can be cast in the diffusion region as $E^z\sim v_1B_1\sim v_2B_2.$ Then, from Eq. \eqref{eq-ohmg3} we get
\begin{align}
    \frac{v_1}{c}&\sim \frac{1}{\gamma_{\text{out}}}\frac{L}{\delta}\left[f_\text{out}\gamma_\text{out}\frac{v_\text{out}}{c}\frac{\lambda^2_e}{4L^2}+\frac{1}{S} \right]\ ,\label{v1_out}
\end{align}
where $S=4\pi L c/\eta$ is the relativistic Lundquist number.
Since $v_1B_1\sim v_2B_2$, it is straightforward to show that an analogue expression for $v_2/c$ can be found.

On the other hand, Eq.~\eqref{eq_cont} allows us to estimate the length rate in the reconnection layer as
\begin{align}
    \frac{L}{\delta}&\sim \frac{2\rho_{\text{out}}\gamma_{\text{out}}v_{\text{out}}}{\rho_1\gamma_1 v_1 + \rho_2 \gamma_2 v_2} \label{length_rate} \ .
\end{align}
Then, using Eq. \eqref{length_rate}, altogether with Eq. \eqref{v1_out} multiplied by $\rho_1$, and its analogue equivalent for $v_2 B_2$ multiplied by $\rho_2$, we find that the sum of both expressions becomes finally
\begin{align}
   \big(&\rho_1\gamma_1 v_1 + \rho_2 \gamma_2 v_2\big)\left(\rho_1 \frac{v_1}{c}+\rho_2\frac{v_2}{c}\right) \nonumber \\
   &\sim 2\rho_{\text{out}}\left(\rho_1+\rho_2\right)v_{\text{out}}\left[ f_\text{out}\frac{\gamma_\text{out} v_\text{out}}{c}\frac{\lambda^2_e}{4L^2}+\frac{1}{S} \right]\, . \label{final_rel}
\end{align}

This is the main equation
of this work. From here, we can study different reconnection rates for diverse asymmetric plasma conditions. Two important cases are studied in the following.

\subsection{Asymmetric inflow velocities and magnetization}\label{asy1sec}

In this case, we can assume that the  magnetic reconnection occurs due to the unbalance in the inflow plasma velocities and its magnetization. In such case, let us consider mildly-relativistic inflow velocities $v_1\neq v_2$, but with $\gamma_1 \sim 1$ and $\gamma_2\sim 1$.  Then, if this is the only source of inflow asymmetry, then we can take
$\rho_1\sim\rho_2\sim\rho_{\text{in}}$. In this way, Eq. \eqref{final_rel} reduces to 
\begin{align}
	\frac{\left(v_1+v_2\right)^2}{4v^2_{\text{out}}} \sim \gamma_\text{out}\frac{\rho_{\text{out}}}{\rho_{\text{in}}}\left[f_\text{out}\frac{\lambda^2_e}{4L^2}+\frac{1}{S}\frac{c}{\gamma_\text{out}v_\text{out}}\right]\ .  \label{R_1pp}
\end{align}
Eq. \eqref{R_1pp} suggests that the reconnection rate can be defined as $R_1=\left(v_1+v_2\right)/2v_{\text{out}}$, in order to consider the inflow asymmetry. Therefore, using Eqs. \eqref{gamma} and \eqref{R_1}, we have  
\begin{align}
	R_1 \sim \Bigg[\frac{\rho_{\text{out}}}{\rho_{\text{in}}}\Bigg(f_\text{out}\frac{\lambda^2_e}{4L^2}&\sqrt{1+\frac{\rho_\text{in}\left(\sigma_1+\sigma_2\right)}{2 \rho_\text{out}f_\text{out}}} \nonumber \\
	&+\frac{1}{S}\sqrt{1+\frac{2 \rho_\text{out}f_\text{out}}{\rho_\text{in}\left(\sigma_1+\sigma_2\right)}}\Bigg)\Bigg]^{1/2}\ .\label{R_1p}
\end{align}
where we  have $\rho_{\text{out}}\neq {\rho_{\text{in}}}$, in general. If we consider that $\rho_{\text{out}}= {\rho_{\text{in}}}$, the above reconnection rate becomes
\begin{align}
	R_1 \sim \Bigg(f_\text{out}\frac{\lambda^2_e}{4L^2}&\sqrt{1+\frac{\sigma_1+\sigma_2}{2 f_\text{out}}} +\frac{1}{S}\sqrt{1+\frac{2 f_\text{out}}{\sigma_1+\sigma_2}}\Bigg)^{1/2}\ . \label{R_1}
\end{align}
The asymmetry in the magnetization has an impact in the reconnection rate, which can be enhanced by thermal-inertial effects, proportional to $\lambda_e^2$. On the other hand, in the low magnetization limit, it is possible to show that the above procedure allow us to recover the results of Ref.~\cite{asenjocomisso2014}.

\subsection{Asymmetric plasma densities and symmetric inflow velocities}
\label{asy2sec}

In this case, we can consider an asymmetric magnetic reconnection scenario where the inflow velocities are mainly symmetric $v_1\sim v_2 \sim v_{\text{in}}$, but the inflow plasma densities are different. 
By Ohm's law, this imply that $B_1\sim B_2$, and therefore $\sigma_1\rho_1\sim\sigma_2\rho_2$.
In this case, from Eq.~\eqref{final_rel}, we get
\begin{align}
	 \left(\rho_1+ \rho_2 \right)\frac{v^2_{\text{in}}}{c} & \sim 2\rho_{\text{out}}v_{\text{out}}\left[ f_\text{out}\frac{\gamma_\text{out} v_\text{out}}{c}\frac{\lambda^2_e}{4L^2}+\frac{1}{S} \right]\ , \label{final_2}
\end{align}
suggesting that the outflow density can be considered to be $\rho_{\text{out}}\sim \left(\rho_1+\rho_2\right)/2$, and that the
reconnection rate can be defined as $R_2=v_{\text{in}}/v_{\text{out}}.$ Thus, from Eqs. \eqref{gamma} and \eqref{final_2} we obtain
\begin{align}
	 R_2\sim &\sqrt{\gamma_{\text{out}}\left[ f_\text{out}\frac{\lambda^2_e}{4L^2}+\frac{1}{S}\frac{c}{\gamma_{\text{out}}v_{\text{out}}} \right]}\   \nonumber \\
	   \sim & \left[ f_\text{out}\frac{\lambda^2_e}{4L^2}\sqrt{1+\frac{2\sigma_1\sigma_2}{f_\text{out}\left(\sigma_1+\sigma_2\right)}}\right.\nonumber \\ 
	 &\qquad\qquad\qquad  \left.+\frac{1}{S}\sqrt{1+\frac{f_\text{out}\left(\sigma_1+\sigma_2\right)}{2\sigma_1\sigma_2}} \right]^{1/2} \ . \label{R_2}
\end{align}
Anew, thermal-inertial effects enhance the effect of magnetization in the reconnection.
Notice how different is this reconnection rate in comparison with the previous one \eqref{R_1}.

Lastly, it is illustrative to show how this result reduces to known cold asymmetric limits, when $f_\text{out}=1$. In this case  Eq.~\eqref{gamma} 
becomes
\begin{equation}
    \gamma_\text{out}=\sqrt{1+\frac{2\sigma_1\sigma_2}{\sigma_1+\sigma_2}}\, .
\end{equation}
On the other hand, combining Eqs.~\eqref{v1_out} and \eqref{R_2} we get
$R_2=\sqrt{\gamma_\text{out}{\delta} R_2/L}$.
From these equations, we finally obtain
\begin{equation}
    R_2=\frac{\delta}{L}\sqrt{1+\frac{2\sigma_1\sigma_2}{\sigma_1+\sigma_2}}\, ,
    \label{coldmbarek}
\end{equation}
thus recovering the results presented in Refs.~ \cite{Lyubarsky2005,asenjocomisso2014,mbarek2022}.

\section{Remarks on the two reconnection rates}\label{discussion}

The reconnection rates presented in Eqs. \eqref{R_1} and \eqref{R_2} show a generalization of relativistic asymmetric magnetic reconnection with thermal-inertial effects. 
Our results exhibit that it is possible to obtain two different reconnection rates depending on the asymmetric inflow conditions of the reconnecting plasma.

In Sec.\ref{asy1sec}, to estimate the magnetic reconnection rate $R_1$, we consider two different regions of plasma flowing into the diffusion region with different velocities, from above and below the $x$ axis, $v_1$ and $v_2$, respectively. The magnetic field strength is also asymmmetric for plasma inflowing from above $B_1$ and below $B_2$ the reconnection region. Each plasma has its own density, but we consider the simplest case when they are approximately equal. Still, this density is different for the outflow plasma. 
These conditions make possible to define $R_1$ as  in Eq.~\eqref{R_1p}, representing a generalization of results seen in Ref.~\cite{mbarek2022}.
This reconnection rate shows how thermal-inertial effects and relativistic effects increase the value of $R_1$ in an asymmetric context. Thermal-inertial effects are present through the electron skin depth $\lambda_e$ and the magnetic diffusion region length $L$ ratio. In general, the quantity $\lambda_e/L \ll 1$ because of how the reconnection must occur in the diffusion region.
Relativistic effects are present through enthalpy $h_{\text{out}}$ or the thermal function $f_{\text{out}}$, that can be described by the particles’ rest mass and the relativistic temperature ratio, taking into account the asymmetric inflow conditions. Generally, $f_{\text{out}}\geq 1$, where $f_{\text{out}}=1$ represents a nonrelativistic plasma. Depending on $f_{\text{out}}$, this thermal-inertial term $f_{\text{out}}\lambda^2_e/4L^2$ on Eq. \eqref{R_1p}, always grows.  
In this way, the expression for $R_1$ exhibit that thermal-inertial effects increase the reconnection rates and the asymmetric effects in the reconnection rate can be increased by high-temperatures ($f_\text{out}\gg 1$) in plasmas, compared to the purely resistive case \cite{Lyubarsky2005}. 

Contrastingly, in Sec.\ref{asy2sec}, we studied a different asymmetric case, by choosing $v_1\sim v_2 = v_{\text{in}}$, but considering that each plasma maintains its inflow density from above and below the reconnection region, $\rho_1$ and $\rho_2$, respectively. This suggests that the density for the outflow plasma can be considered as $\left(\rho_1 + \rho_2\right)/2$. Using these conditions in Eq. \eqref{final_rel}, the reconnection rate $R_2$ is obtained  in Eq.~\eqref{R_2}. 
This case exhibits that thermal-inertial effects and relativistic effects keep on increasing the reconnection rate, even with different asymmetric conditions. 

\begin{figure}[h!]
    \centering
    \subfloat[\centering Ratio $R_1/R_2$ for $f \lambda_e^2 S/(4 L^2)=1/500$.]{{\includegraphics[width=8.4cm]{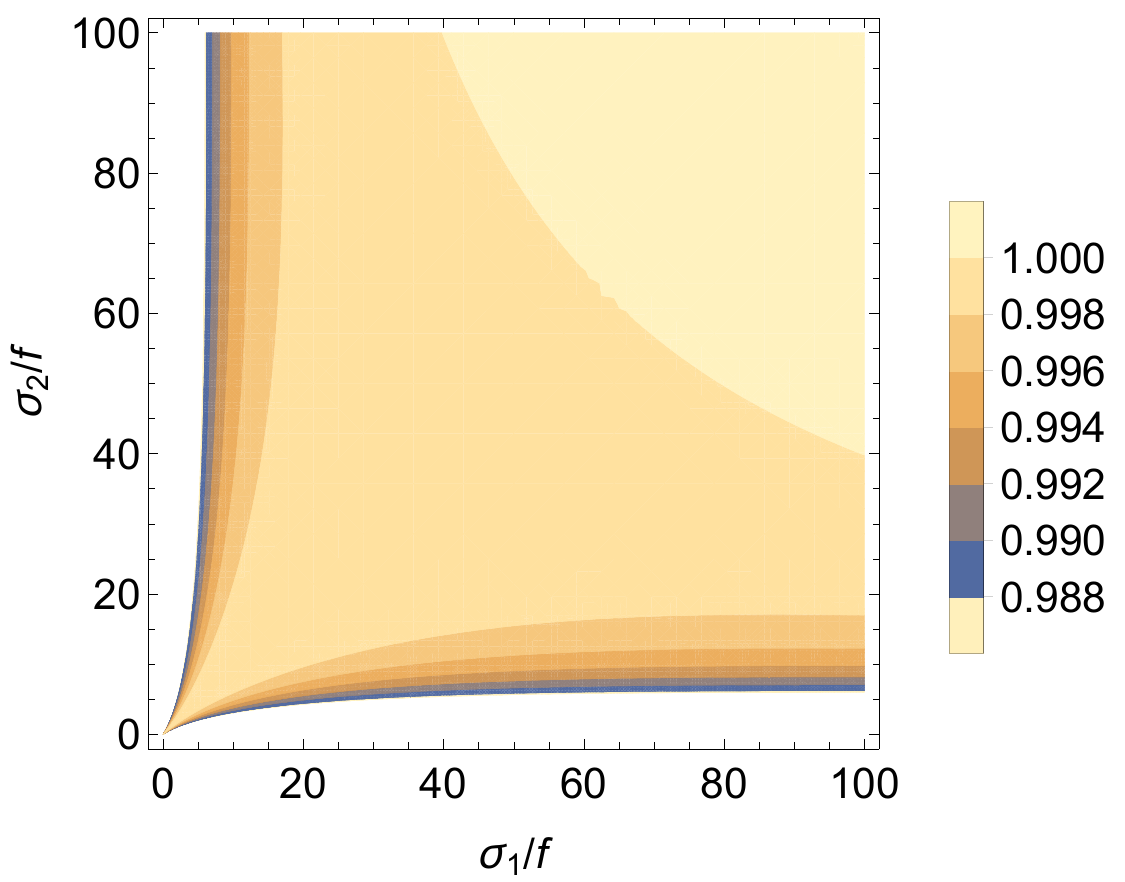} }}%
    \qquad
    \subfloat[\centering  Ratio $R_1/R_2$ for $f \lambda_e^2 S/(4 L^2)=1/50$.]{{\includegraphics[width=8.4cm]{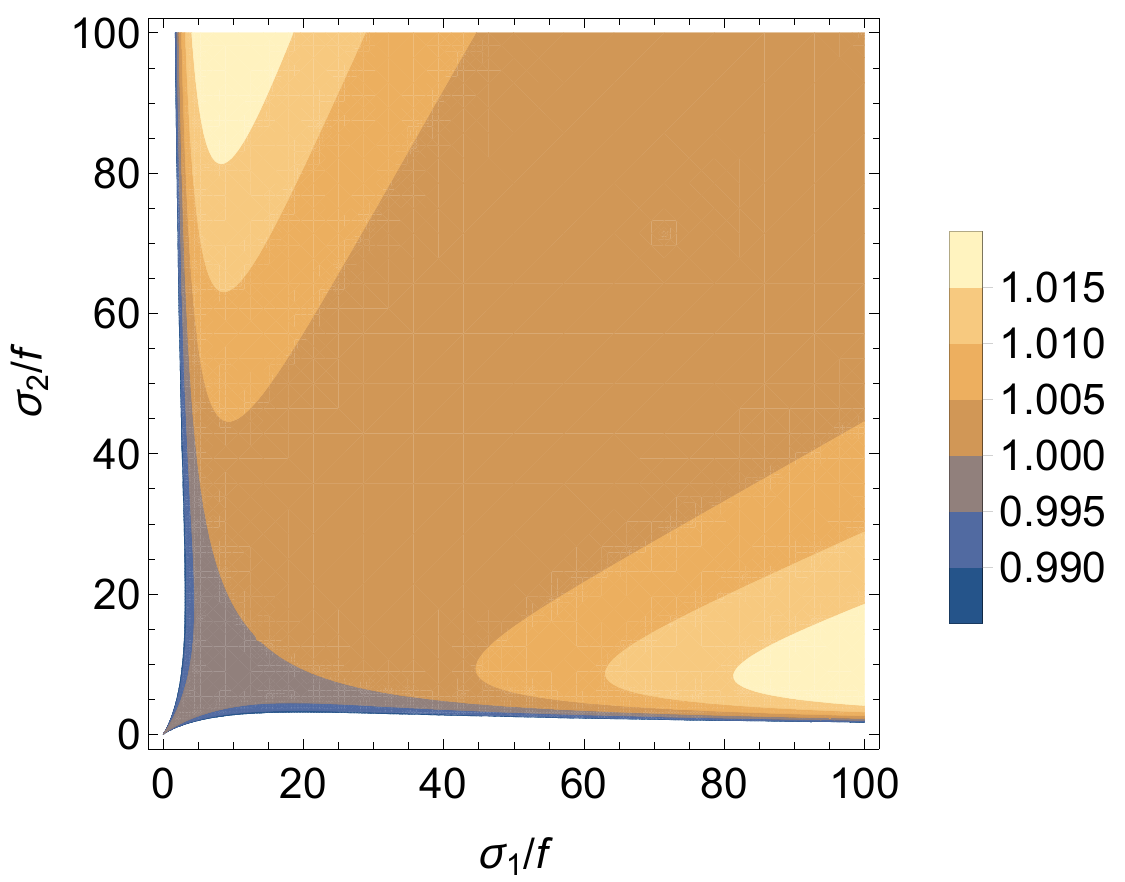} }}%
    \qquad
    \subfloat[\centering  Ratio $R_1/R_2$ for $f \lambda_e^2 S/(4 L^2)=1/2$.]{{\includegraphics[width=8.4cm]{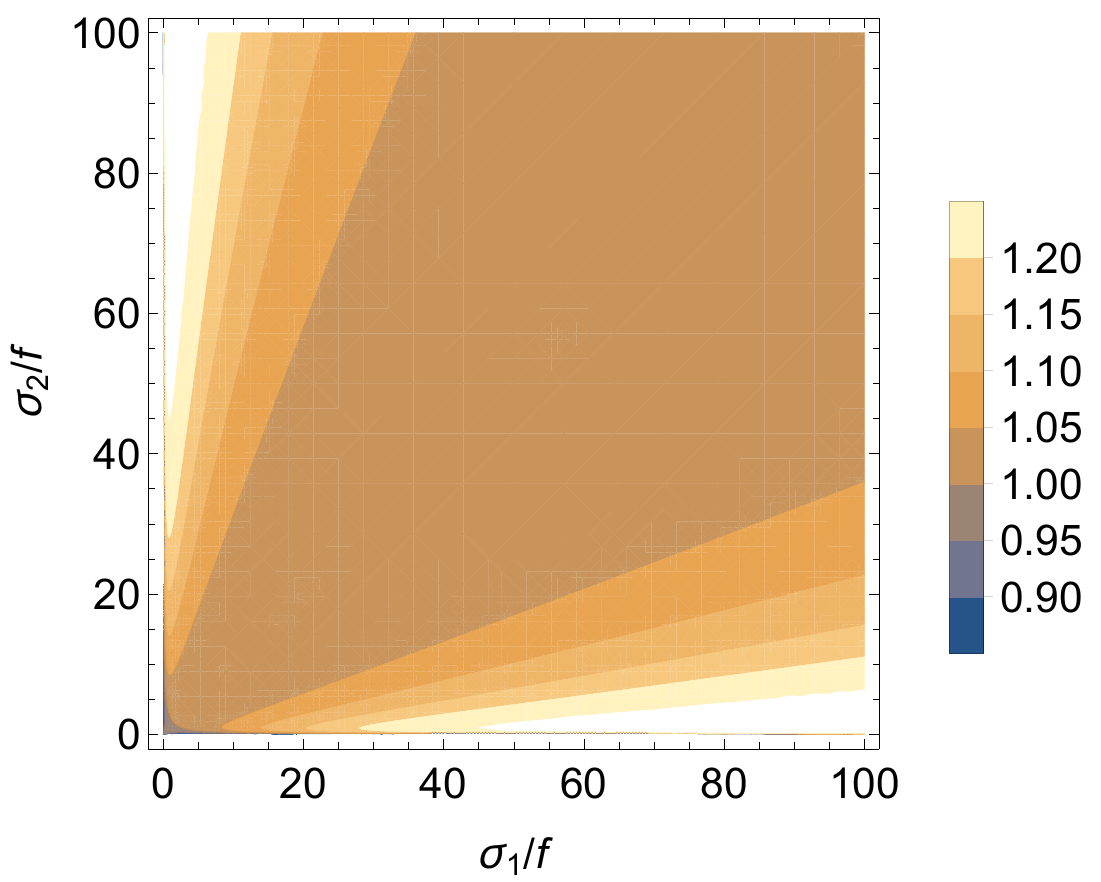} }}%
    \caption{Contour plots for the ratio $R_1/R_2$ [from reconnection rates \eqref{R_1} and \eqref{R_2}] for three different strength of thermal-inertial effects. Notice how $R_1/R_2<1$ or $R_1/R_2>1$ depending on magnetization and thermal-inertial effects.}
    \label{fig22}
\end{figure}

These two reconnection rates, $R_1$ and $R_2$, have opposite behaviors on the resistivity plasma contribution part (proportional to $1/S$) and  on thermal-inertial plasma contribution part (proportional to $\lambda_e^2$). And this is only due to the asymmetry in the inflow plasma magnetization.
Because of $(\sigma_1+\sigma_2)/(2\sigma_1\sigma_2)\geq 2/(\sigma_1+\sigma_2)$ for any possible magnetizations, then
we found that the resistivity part of the reconnection rate $R_2$ is greater than the resistivity part of reconnection rate $R_1$. On the contrary, the thermal-inertial part of the reconnection rate $R_2$ is less than the analogous part of reconnection rate $R_1$. Thus, the different asymmetric plasma configurations  produces different reconnection outcomes. In general, the dependence on the asymmetry of magnetization, and on the strength value of the thermal-inertial effects, establish which reconnection rate is larger. This is exemplified in Fig.~\ref{fig22}. The ratio $R_1/R_2$ is plotted in terms of $\sigma_1/f$ and $\sigma_2/f$, where the reconnection rates are defined in \eqref{R_1} and \eqref{R_2}, respectively. In the plots, we show the contours for different values of the ratio $R_1/R_2$. We present three cases, for $f \lambda_e^2 S/(4 L^2)=1/500$ in Fig.~\ref{fig22}(a),  $f \lambda_e^2 S/(4 L^2)=1/50$ in Fig.~\ref{fig22}(b), and  $f \lambda_e^2 S/(4 L^2)=1/2$ in Fig.~\ref{fig22}(c). 
From these three plots, it is clear that $R_1/R_2$ could be less or larger than unity depending on asymmetry on magnetization and on how important are thermal-inertial effects in plasma dynamics. Larger asymmetric magnetizations in the inflow plasmas (for example, $\sigma_1\gg \sigma_2$) allow to have (in general) larger reconnection rate $R_1$ over $R_2$. Furthermore, as the thermal-inertial effects become relevant, the reconnection rate $R_1$ becomes dominant for a larger range of values of the asymmetric magnetizations $\sigma_1$ and $\sigma_2$.

This issue is important, because of recent studies on  asymmetric  relativistic reconnection have only considered the cold limit \eqref{coldmbarek} of the  rate $R_2$ as the outcome of the magnetic reconnection process \cite{mbarek2022}.  In the simplest case when thermal-inertial effects are neglected, then $R_1/R_2<1$ always, implying the importance of $R_2$ in this limit.
However, from the previous analysis, we infer that when relativistic effects become
 very important, with larger asymmetric magnetizations and larger thermal-inertial effects, $R_1$ (instead of $R_2$) is the most relevant  rate for the reconnecting dynamics. For instance, for the extreme case of $\sigma_1=100$, $\sigma_2=1$, and $f \lambda_e^2 S/(4 L^2)=0.55$, we find that $R_1$ is about 50\% larger than $R_2$.

Therefore,  the study on how the asymmetric properties of the inflow plasma affect the reconnection rate, and how its relevant relativistic features determine their strength, is an issue that must be addressed in future investigations.
We believe
that the recognition of the importance of these two points on the definition of diverse kinds of asymmetric relativistic reconnection rates is the fundamental  result of this work.



\end{document}